# A probabilistic graphical model approach in 30 m land cover mapping with multiple data sources

Jie Wang [a], Luyan Ji [b], Xiaomeng Huang [b], Haohuan Fu [b,*], Shiming Xu [b], Congcong Li [b]

[a] *State Key Laboratory of Remote Sensing Science, Institute of Remote Sensing and Digital Earth, Chinese Academy of Sciences, Beijing, 100101, China;*

[b] *Ministry of Education Key Laboratory for Earth System Modeling, Center for Earth System Science, Tsinghua University, Beijing, 100084, China;*

**\*** Corresponding author. E-Mail: haohuan@tsinghua.edu.cn

# A probabilistic graphical model approach in 30 m land cover mapping with multiple data sources

There is a trend to acquire high accuracy land-cover maps using multi-source classification methods, most of which are based on data fusion, especially pixel- or feature-level fusions. A probabilistic graphical model (PGM) approach is proposed in this research for 30 m resolution land-cover mapping with multi-temporal Landsat and MODerate Resolution Imaging Spectroradiometer (MODIS) data. Independent classifiers were applied to two single-date Landsat 8 scenes and the MODIS time-series data, respectively, for probability estimation. A PGM was created for each pixel in Landsat 8 data. Conditional probability distributions were computed based on data quality and reliability by using information selectively. Using the administrative territory of Beijing City (Area-1) and a coastal region of Shandong province, China (Area-2) as study areas, multiple land-cover maps were generated for comparison. Quantitative results show the effectiveness of the proposed method. Overall accuracies promoted from 74.0% (maps acquired from single-temporal Landsat images) to 81.8% (output of the PGM) for Area-1. Improvements can also be seen when using MODIS data and only a single-temporal Landsat image as input (overall accuracy: 78.4% versus 74.0% for Area-1, and 86.8% versus 83.0% for Area-2). Information from MODIS data did not help much when the PGM was applied to cloud free regions of. One of the advantages of the proposed method is that it can be applied where multi-temporal data cannot be simply stacked as a multi-layered image.

1. Introduction

The medium resolution and the free availability of Landsat data make it feasible for use in large-area land-cover mapping. The national land cover database (NLCD 2001, NLCD2006) for the conterminous United States (Homer et al. 2004; Fry et al. 2011), the circa 2010 forest map for China (Li, Wang, Hu, et al. 2014), the first 30 m global land cover maps (Gong et al. 2013), the global tree cover mapping (Hansen et al. 2013; Sexton et al. 2013) and many other large area maps have been derived from Landsat data. However, with a 16-day revisit interval, high-quality Landsat data are not always available for all areas. For example, scenes from Landsat 7 are on average about 35% cloud covered at the global scale (Roy et al. 2008). Growing season data are more useful for land-cover mapping than imagery captured during other periods (Hansen et al. 2013), but clouds are also abundant in many growing season images. To further improve classification results, multi-temporal Landsat data are preferable for land cover classification (Guerschman et al. 2003; Hansen et al. 2013). In circumstances where Landsat data are insufficient, it is also helpful to include other sources, such as the MODerate Resolution Imaging Spectroradiometer (MODIS) time series, as auxiliary data (Hansen et al. 2008; Potapov et al. 2008; Lu et al. 2011; Yu , Wang, and Gong 2013; Li, Wang, Hu, et al. 2014; Chen et al. 2015).

In conventional supervised image classification, samples are usually collected locally when mapping local land covers. Usually only one classifier is trained and applied, with all spectral bands in combination with other features extracted from multi-temporal Landsat data as input. Satisfactory results may not necessarily be obtained when mapping large-area land cover in a similar way. If a corresponding classifier was to be trained for each individual Landsat scene, a large volume of samples would be needed for large-area land-cover mapping. Although 91,433 training sample units were collected in Gong et al. (2013), there were only 10-20 sample units in each Landsat scene. The number of sample units in an individual Landsat scene is scarcely sufficient for training a multi-class classifier, even for the largest sample set provided by Geo-Wiki (Fritz et al. 2012). This can partly be solved by collecting training sample units from spatio-temporal neighbourhood scenes (Gong et al. 2013). Another issue arises due to the sparsity of sample points when stacking multi-temporal Landsat as a multi-layered image for large-area land-cover mapping. When training a supervised classifier on these multi-layered images, the number of features of all the images should be equal, and the sublayers with the same sequence number in different images are better acquired at similar dates, but clear Landsat scenes are rare in many areas (Helmer et al. 2010), and therefore, the conditions are difficult to meet. The Landsat time series cannot be simply treated in a similar way to regular MODIS time series. We term this a temporal inconsistency problem. One possible way to partly overcome the temporal inconsistency problem is to extract features by computing time series metrics, such as minimum, maximum and mean reflectance values (Hansen et al. 2013).

Data fusion is an effective way to integrate multi-source data for large-area land-cover mapping. Most of the pixel-level fusion methods (Gao et al. 2006; Roy et al. 2008; Zhu et al. 2010; Zhang et al. 2013; Walker, de Beurs, and Wynne 2014; Weng, Fu, and Gao 2014) are data-specific. Most of the feature-level fusion methods (Hansen et al. 2008; Potapov et al. 2008; Xin et al. 2013; Lu et al. 2011) are knowledge-based. When using pixel- or feature-level fusion methods, careful data selection is needed for accurate land-cover mapping (Senf et al. 2015). Methods based on a combination of probabilistic or fuzzy decisions (Mangai et al. 2010; Liu, Kelly, and Gong 2006; Liu and Cai 2012; Wang, Zhao, Li, et al. 2015; Solberg, Taxt, and Jain 1996) can also be used for multi-source data classification. In most of these methods, initial decisions were all treated equally, hence useful information as well as errors were incorporated without distinction. The adaptively weighted decision fusion method (Wang, Li, and Gong 2015) can use information selectively. Few errors were introduced when improving the initial classification results. However, a complicated, recursive process is needed when fusing more than two types of decision. It is difficult to incorporate knowledge-based rules into the process.

The probabilistic framework has a central role in scientific data analysis, machine learning and artificial intelligence (Ghahramani 2015). The probabilistic graphical model (PGM) is the dominant paradigm in machine learning for composing multiple probability distributions into a complex model (Ghahramani 2015). The PGMs naturally accommodate the need to fuse multiple sources of information (Jordan 2004). Methods based on undirected graphical models (Liu, Kelly, and Gong 2006; Liu and Cai 2012; Wang, Zhao, Li, et al. 2015; Solberg, Taxt, and Jain 1996) are the only PGM-based land cover mapping methods of which we are aware. However, these models were used only for modelling neighbourhood consistency. In this study, a method based on a directed PGM is proposed for 30 m resolution land-cover mapping with Landsat and MODIS data. In our approach, classifiers were only applied to each single-date Landsat image, instead of using the Landsat image series as input. The initial probability estimation results, as well as the probabilities estimated from the MODIS data were then fused using a PGM. Information from different sources was utilized selectively based on their reliability. The method can be applied to large-area land-cover mapping with temporally inconsistent multi-temporal Landsat images.

## 2. Method

### *2.1 Probabilistic graphical model development*

Directed and undirected PGMs are the two most common forms of PGMs. An undirected PGM is defined based on an undirected graph, and a directed PGM is defined based on a directed graph. The nodes of a graph represent variables of the model, and the edges represent probabilistic interactions among the variables. Undirected PGMs, such as Markov Random Field models (Bishop 2006), are often used for modelling neighbourhood relations. Directed PGMs are often used for modelling causal relations.

The diagram of the proposed PGM is shown in Figure 1. The plate (the rectangle) in the diagram denotes that the subgraph inside the plate repeats $N$ times. This can also be explained as that one node outside of the plate corresponds to $N$ nodes in the plate. Nodes in the graph are: the initial estimation of a MODIS pixel ($M$), the initial estimation of a Landsat-A pixel ($LA$), the initial estimation of a Landsat-B pixel ($LB$), the combined result of Landsat-A and Landsat-B ($LL$), and the estimation for real land cover ($R$). We denoted the random variables of the nodes as: $M$, $LA_n$, $LB_n$, $LL_n$, and $R_n$, where $n \in \{1, \cdots, N\}$.

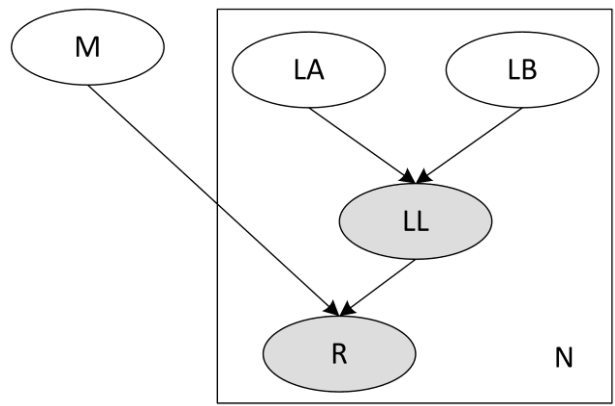

Figure 1. Diagram of the proposed directed probabilistic graphical model. Details are explained in the text.

Events of the random variables are predefined land-cover classes. The joint probability distribution of the variables can be expressed as:

$$P(M, LA_1, LB_1, LL_1, R_1, \cdots, LA_N, LB_N, LL_N, R_N)$$
$$= P(M)P(LA_1)P(LB_1)P(LL_1 | LA_1, LB_1)P(R_1 | LL_1, M) \quad (1)$$
$$\cdots P(LA_N)P(LB_N)P(LL_N | LA_N, LB_N)P(R_N | LL_N, M)$$

where the conditional probability distributions (CPDs) $P(LL_n | LA_n, LB_n)$ and $P(R_n | LL_n, M)$ can be defined based on knowledge (see section 2.2). The most probable land cover class is found by a marginal maximum a posterior (marginal MAP) query (Koller and Friedman 2009) on the proposed PGM:

$$\text{MAP}(R_n) = \underset{c}{\text{argmax}} \, P(R_n = c) \quad (2)$$

where $c$ is a land cover class, and

$$P(R_n) = \sum_{\text{all variables other than } R_n} P(M, LA_1, LB_1, LL_1, R_1, \cdots, LA_N, LB_N, LL_N, R_N) \quad (3)$$

We have

$$\sum_{LA_n, LB_n, LL_n, R_n} P(LA_n)P(LB_n)P(LL_n|LA_n, LB_n)P(R_n|LL_n, M)$$
$$= \sum_{LL_n, R_n} P(R_n|LL_n, M) \sum_{LA_n, LB_n} P(LA_n)P(LB_n)P(LL_n|LA_n, LB_n)$$
$$= \sum_{LL_n, R_n} P(R_n|LL_n, M) \sum_{LA_n, LB_n} P(LA_n, LB_n, LL_n) \quad (4)$$
$$= \sum_{LL_n, R_n} P(R_n|LL_n, M) P(LL_n)$$
$$= \sum_{LL_n, R_n} P(R_n, LL_n | M)$$
$$= 1$$

Therefore

$$P(R_n) = \sum_{\text{all variables other than } R_n} P(M, LA_1, LB_1, LL_1, R_1, \cdots, LA_N, LB_N, LL_N, R_N)$$

$$= \sum_{\text{all variables other than } R_n} P(M) P(LA_1) P(LB_1) P(LL_1|LA_1, LB_1) P(R_1|LL_1, M)$$

$$\cdots P(LA_N) P(LB_N) P(LL_N|LA_N, LB_N) P(R_N | LL_N, M)$$

$$= \sum_{M, LA_n, LB_n, LL_n?} P(M) P(LA_n) P(LB_n) P(LL_n|LA_n, LB_n) P(R_n|LL_n, M)$$

$$\left( \sum_{LA_1, LB_1, LL_1, R_1} P(LA_1) P(LB_1) P(LL_1|LA_1, LB_1) P(R_1|LL_1, M) \right)$$

$$\cdots \left( \sum_{LA_N, LB_N, LL_N, R_N} P(LA_N) P(LB_N) P(LL_N|LA_N, LB_N) P(R_N|LL_N, M) \right)$$

$$= \sum_{M, LA_n, LB_n, LL_n?} P(M) P(LA_n) P(LB_n) P(LL_n|LA_n, LB_n) P(R_n|LL_n, M) \quad (5)$$

The marginal probability distribution $P(R_n)$ can be computed using a two-step variable elimination shown in (6) and (7):

$$P(LL_n) = \sum_{LA_n, LB_n} P(LA_n) P(LB_n) P(LL_n|LA_n, LB_n) \quad (6)$$

$$P(R_n) = \sum_{M, LL_n} P(M) P(R_n|LL_n, M) P(LL_n) \quad (7)$$

Therefore, the land-cover mapping result can also be obtained by a two-step combination. The first step is only performed between Landsat data, and the second step is performed between the first-step result and the MODIS data. The first step can be simplified as $P(LL_n) = P(LA_n)$ when only Landsat-A is used.

The variable elimination algorithm has exponential time complexity (Koller and Friedman 2009). It can be estimated that the computational complexity of (6) and (7) is $O(C^3)$, where $C$ is the number of land cover classes. Thus the computational complexity is $O(C^3N)$ when performing the algorithm on all the $N$ pixels. The complexity can be reduced to $O(CN)$ if most values of the CPDs are zero, as described in section 2.2.

*2.2 Conditional probability distributions*

Conditional probability distributions (CPD)s can be determined by reference to data quality, local land-cover knowledge, and auxiliary information. For simplicity, the CPDs $P(LL_n | LA_n, LB_n)$ and $P(R_n | LL_n, M)$ used in this study were defined based on data quality only. Suppose that Landsat-A is cloud-free, and Landsat-B is partly covered by cloud and cloud shadow. Thus, fractions of cloud, soil, vegetation and dark objects were calculated based on the linear mixing model with the abundance sum-to-one constraint. The corresponding endmembers were automatically extracted using the modified N-FINDR (Ji et al. 2015). We set $f_n = (c+s)/(c+s+v+d)$, where $c, s, v$ and $d$ are the fractions of cloud, soil, vegetation and dark objects respectively. The CPD $P(LL_n | LA_n, LB_n)$ was defined as:

$$P(LL_n | LA_n, LB_n) = \begin{cases} 1, & \text{if } LL_n = LA_n = LB_n \\ 1 - 0.5(1 - f_n), & \text{if } LL_n = LA_n \neq LB_n \\ 0.5(1 - f_n), & \text{if } LL_n = LB_n \neq LA_n \\ 0, & \text{if } LL_n \neq LA_n, LL_n \neq LB_n \end{cases} \quad (8)$$

This definition means that probabilities of $LL_n$ are more similar to probabilities of $LA_n$ than probabilities of $LB_n$ when $f_n$ is large. The extreme cases are: $P(LL_n) = P(LA_n)$ when $f_n = 1$, and $P(LL_n) = \frac{1}{2}[P(LA_n) + P(LB_n)]$ when $f_n = 0$.

The Landsat data were reprojected onto the MODIS sinusoidal grid to find a pixel-to-pixel correspondence. Pixels of Landsat data corresponding to the same MODIS pixel were set together as one group. For a given pixel $LL_n$, class $c = MAP(LL_n) = \underset{c}{\arg\max} \, P(LL_n = c)$ is the most probable class of the pixel. Pixels in the same group are considered similar to $LL_n$ if their most probable classes are also class $c$. We set $w_n = g_n / (g_n + 1 - m_n)$, where $g_n$ is the percentage of pixels in the same group that are similar to $LL_n$; and $m_n$ is the percentage of missing values for the time series of the corresponding MODIS pixel, recorded as described in section 3.2. Then the CPD $P(R_n | LL_n, M)$ was defined as:

$$P(R_n | LL_n, M) = \begin{cases} 1, & \text{if } R_n = LL_n = M \\ 1 - 0.5w_n, & \text{if } R_n = LL_n \neq M \\ 0.5w_n, & \text{if } R_n = M \neq LL_n \\ 0, & \text{if } R_n \neq LL_n, R_n \neq M \end{cases} \quad (9)$$

The reasons behind the definitions of CPDs $P(LL_n | LA_n, LB_n)$ and $P(R_n | LL_n, M)$ are similar. Information obtained from high-quality data should be heavily weighted. Less credible information should be less heavily weighted.

Most values of the CPDs defined in Eqs (8) and (9) are zeroes. Thus, the variable elimination procedure (6) can be simplified as:

$$\begin{aligned} P(LL_n &= c) \\ &= P(LA_n = c)P(LB_n = c) \\ &\quad + P(LA_n = c)P(LB_n \neq c)P(LL_n = c | LA_n = c, LB_n \neq c) \\ &\quad + P(LA_n \neq c)P(LB_n = c)P(LL_n = c | LA_n \neq c, LB_n = c) \\ &= P(LA_n = c)P(LB_n = c) \\ &\quad + P(LA_n = c)[1 - P(LB_n = c)][1 - 0.5(1 - f_n)] \\ &\quad + [1 - P(LA_n = c)]P(LB_n = c)[0.5(1 - f_n)] \end{aligned} \quad (10)$$

where $c$ is a land cover class. The computational complexity is $O(CN)$ when applying Eq. (10) to all the $N$ pixels, where $C$ is the number of land cover classes. The procedure (7) can be simplified in a similar way. Thus, the computational complexity of the variable elimination procedure is far lower than that for the Support Vector Machine (SVM) algorithm (Chang and Lin 2011).

## 3. Experiments

### *3.1 Study areas*

Two study areas are selected, the capital of China, Beijing (referred to as Area-1), and a coastal region of Shandong province, China (referred to as Area-2) (Figure 2). Area-1 is located in the northeast edge of the North China Plain, between latitudes 39°26′N and 41°03′N, and longitudes 115°25′E and 117°30′E. Impervious layer, cropland, forest and shrubland are the four major land cover classes in this area. Impervious layer and cropland are distributed mainly in the plain areas. Wheat, corn and soybean are the major agricultural crops. Forest and shrubland are distributed mainly in the mountain areas and most trees and shrubs are deciduous. Cropland, forest, waterbodies and impervious layer are the four major land cover classes in Area-2.

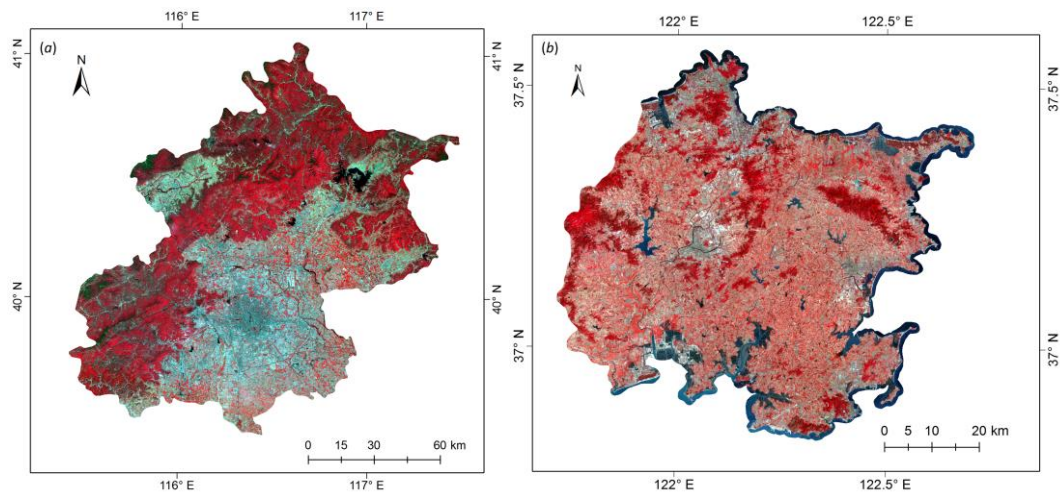

Figure 2. Study areas (data source: Landsat 8, 12 May 2013; Landsat 8, 16 May 2013, showing in RGB=543 OLI band composition).

*3.2 Data and preprocessing*

Five Landsat 8 images (path/row: 123/32 and 123/33, acquired on 12 May 2013; path/row 123/32 and 123/33, acquired on 1 September 2013; path/row 119/34, acquired on 16 May 2013) and three tiles (h26v04, h26v05 and h27v05) of MODIS (MOD13Q1) time series for the year 2013 were downloaded from the United States Geological Survey (USGS) website (Landsat, http://earthexplorer.usgs.gov; MODIS, http://e4ftl01.cr.usgs.gov). The Landsat 8 images acquired in May are cloud free, while the images acquired in September are partly covered by cloud. The Landsat 8 images were used as the primary source of data. The MOD13Q1 time series were used as auxiliary data.

The Landsat 8 level-1 product contains nine Operational Land Imager (OLI) bands and two Thermal Infrared Sensor (TIRS) bands. The OLI bands include one panchromatic band at 15 m resolution and eight multispectral bands at 30 m resolution. The TIRS bands were collected at 100 m and resampled to 30 m resolution. The Landsat 8 surface reflectance product, which is derived from the level-1 product, contains seven multispectral bands (1-7) and three cloud information bands at 30 m resolution. The seven multispectral bands and the cloud mask (CFMask; Zhu and Woodcock 2012) layer of the surface reflectance product, as well as the panchromatic band and the two TIRS bands of the level-1 product were selected for use. For each pair of images (path/row: 123/32 and 123/33) acquired on the same date, these selected bands were mosaicked individually. The two mosaicked Landsat 8 data sets for Area-1 are referred to herein as Landsat-A1 (acquired on 12 May 2013) and Landsat-B1 (acquired on 1 September 2013). And the Landsat 8 data set for Area-2 (path/row: 119/34) is referred to as Landsat-A2.

MOD13Q1 data in MODIS Collection 5 are provided every 16 days at a nominal 250 m resolution. The MOD13Q1 product contains an enhanced vegetation index (EVI) layer, a normalized difference vegetation index (NDVI) layer, four spectral bands (RED (620–670 nm), near-infrared (NIR) (841–876 nm), BLUE (459–479 nm), and mid-infrared (MIR) (2105–2155 nm)), and other information layers. The BLUE and MIR bands were collected at 500 m and resampled to 250 m resolution. EVI is less sensitive to soil and atmospheric effects than NDVI (Waring et al. 2006). Thus, EVI and the four spectral bands were selected as features for the time series. The cloudy pixels and the fill values of the time series from MOD13Q1 were first set to missing values. The percentage of missing values for each of the time series was recorded. Then, the EVI subseries were smoothed with a Savitzky-Golay-filter-based method (Chen et al. 2004), and the other (spectral bands) subseries were simply smoothed with this filter.

*3.3 Sample collection*

The classification system used for Area-1 is composed of seven classes: cropland, forest, grassland, shrubland, waterbodies, impervious, and barren land. The definitions of the classes are taken from a Landsat-based global land-cover mapping project (Gong et al. 2013), except that bare cropland is defined as belonging to cropland. The classification system used for Area-2 is composed of four classes: cropland, forest, waterbodies, and impervious. Sample class labels were interpreted using high resolution Google Earth images, field survey photos and field knowledge as supplementary information.

Validation samples were collected from randomly distributed positions. Totally 439 and 423 single-pixel points were obtained for Area-1 and Area-2, respectively. Training samples for Landsat 8 data were obtained as small and large sample units. The small sample units were extracted as single pixels, and only used for Landsat 8 data classification. The large sample units were extracted as multiple pixels in Landsat 8 images, and as single-pixels in MODIS data. In Area-1, we acquired 11669 and 193 training sample pixels for Landsat 8 and MODIS data, respectively. And in Area-2, the pixel numbers are 34495 and 244. The spatial distribution of the training and validation samples is shown in Figure 3. All samples have been checked to avoid spatial overlap.

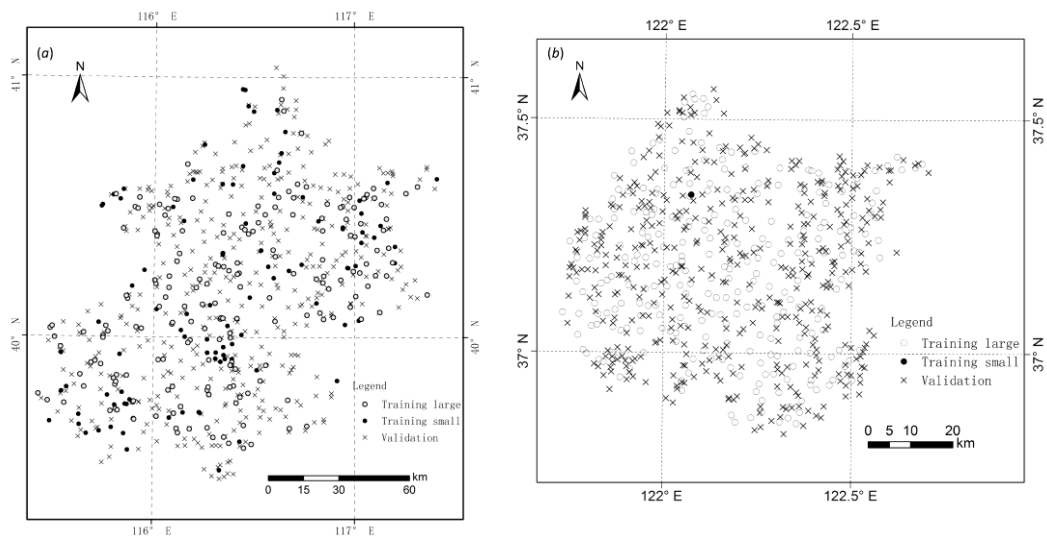

Figure 3. Spatial distributions of training and validation samples.

*3.4 Model implementation*

The block diagram of the implementation is shown in Figure 4. Three independent classifiers were each applied to the Landsat-A, Landsat-B data and the MODIS time series. The predicted probabilities, cloud and shadow fractions for Landsat data and reliabilities for MODIS data were then input into a PGM. Information from Landsat-A and MODIS were used as input when Landsat-B is not included for use. The output of the PGM was the multi-source land-cover mapping result. The Landsat-A1 and Landsat-B1 were used as Landsat-A and Landsat-B respectively when applying this implementation to Area-1. The Landsat-A2 was used as Landsat-A when applying it to Area-2. In this section, we first introduce the classifiers used in this study and the probability estimation, and then describe in detail the implementation of the PGMs for data.

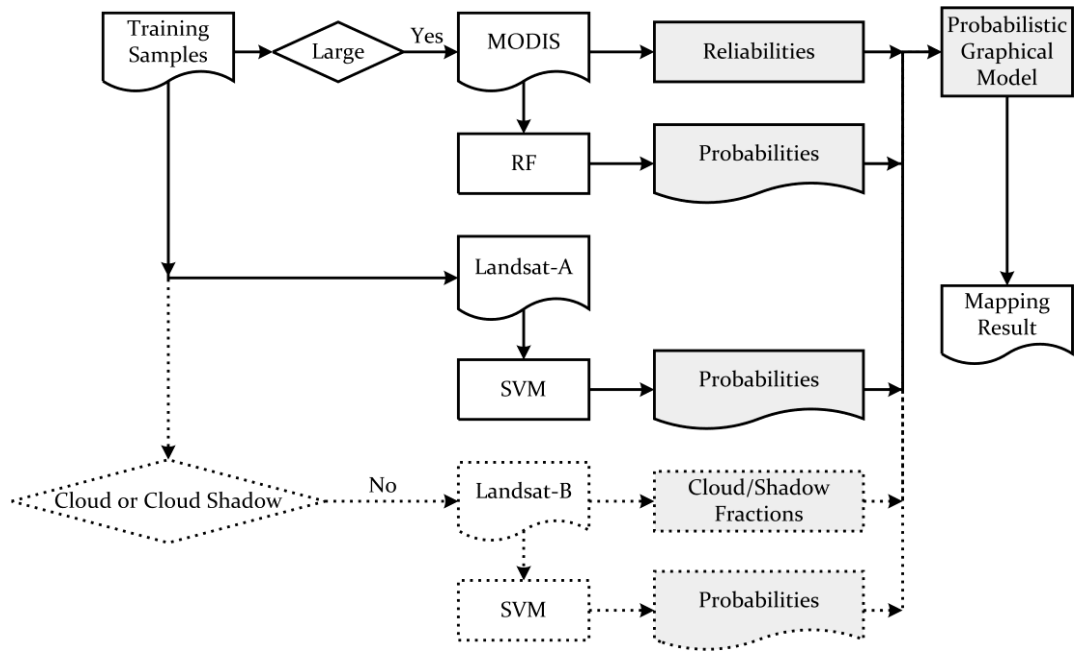

Figure 4. Block diagram of the implementation.

*3.4.1    Support Vector Machine and Random Forests*

Support Vector Machine (SVM; Cortes and Vapnik 1995) and Random Forests (RF; Breiman 2001) are two frequently used supervised learning methods. SVM has proved effective in Landsat-based land-cover classification (Huang, Davis, and Townshend 2002; Liu, Kelly, and Gong 2006; Pal and Mather 2005). RF has proved effective when using multi-seasonal imagery (Rodriguez-Galiano, Chica-Olmo, et al. 2012). Therefore, the SVM classifier with radial basis function kernel (Chang and Lin 2011) was applied to the classification and probability estimation on the Landsat 8 data. And the RF classifier was applied to the probability estimation on the MODIS time series. SVM can perform a non-linear classification by mapping input features into high-dimensional feature spaces using a kernel function. It performs class probability estimation by solving a restricted minimization problem (Chang and Lin 2011; Wu, Lin, and Weng 2004). RF is an ensemble classifier that consists of multiple decision trees. It can maintain accuracy when missing values exist (Rodriguez-Galiano, Ghimire, et al. 2012; Li, Wang, Wang, et al. 2014). The class probability estimation of an RF classifier can be obtained by voting among decision trees.

*3.4.2   Probability estimation*

Two independent SVM classifiers were applied for probability estimation on Landsat-A and Landsat-B respectively. The default one-versus-one strategy of LIBSVM (Chang and Lin 2011) was used for multi-class classification. All collected training samples were used to train the SVM classifier for Landsat-A. And for Landsat-B, samples located in regions declared as cloud or cloud shadow in CFMask were excluded. Input features include seven surface reflectance bands (1-7), two TIRS bands, one NDVI channel and six texture features. NDVI is calculated from the surface reflectance band 4 (Red) and 5 (NIR) as NDVI = (NIR - Red) / (NIR + Red) (Rouse et al. 1974). Three grey-level co-occurrence matrix texture features (Haralick, Shanmuga, and Dinstein 1973; Gong, Marceau, and Howarth 1992; Caridade, Marçal, and Mendonça 2008) including mean, contrast and entropy were extracted from the surface reflectance band 5 (NIR) and the panchromatic band, respectively. The co-occurrence matrix parameters were set as follows: grey levels = 16, window size = 7, and offset = (1, 1). Textures extracted from the panchromatic band were resampled to 30 m resolution. The texture features and parameters were chosen in this study based on experimental findings, and were used as a compromise between land-cover mapping accuracy and computational load. Probability estimation on the MODIS time series was performed using a 200-tree RF classifier. The 193 single-pixel training samples were used to train the RF classifier. Input features are the filtered time series of EVI, RED, NIR, BLUE and MIR.

### 3.4.3 *PGMs for data*

On implementing the proposed method, an independent PGM was created for each pixel in Landsat 8 data. The probabilities estimated for each pixel in MODIS and Landsat data were used as prior distributions. The distributions of Landsat-A and Landsat-B pixels were first combined to acquire probability distributions of the intermediate result *LL*. The $f_n$ value for $P(LL_n | LA_n, LB_n)$ was an approximate estimation of cloud-and-shadow fraction. It was computed only for pixels located in cloud or cloud shadow regions masked by the CFMask layer of Landsat-B, and simply assigned as zeroes for other pixels.

The final result *R* was acquired by combining the probabilities of *LL* and probabilities estimated from MODIS data. This combination was only applied to cloud regions and cloud shadow regions masked by the CFMask layer of Landsat-B. The probability distribution of *LL* was simplified as $P(LL_n) = P(LA_n)$, and the Landsat-MODIS combination was applied to all the pixels of *LL* when only Landsat-A was used.

Although only data quality was considered in the proposed PGM, multiple knowledge-based rules can be added to create a more complicated model. The estimated probability distributions of extra data sources can also be added as new random variables. A large graph structure is needed to model multiple rules as interactions between random variables.

## 4. Results and discussion

The land-cover mapping results (MAP-R1 for Area-1 and MAP-R2 for Area-2) acquired using the proposed method are shown in Figure 5. Extra land-cover maps were generated for Area-1 to compare with MAP-R1. Three maps were created in the process of getting MAP-R1: the map acquired from Landsat-A1 (MAP-A1), from Landsat-B1 (MAP-B1), and from the intermediate result $LL$ (MAP-LL). The map MAP-AM1 was acquired using the proposed method where only Landsat-A1 and MODIS data were used (Landsat-B1 was not included). We also tested the use of stacking Landsat-A1 and Landsat-B1 as a multi-layered image. Two classifiers, SVM and RF, were trained to create two land-cover maps (MAP-AB-SVM and MAP-AB-RF). The land-cover map acquired from Lnadsat-A2 (MAP-A2) was used to compare with MAP-R2 for Area-2.

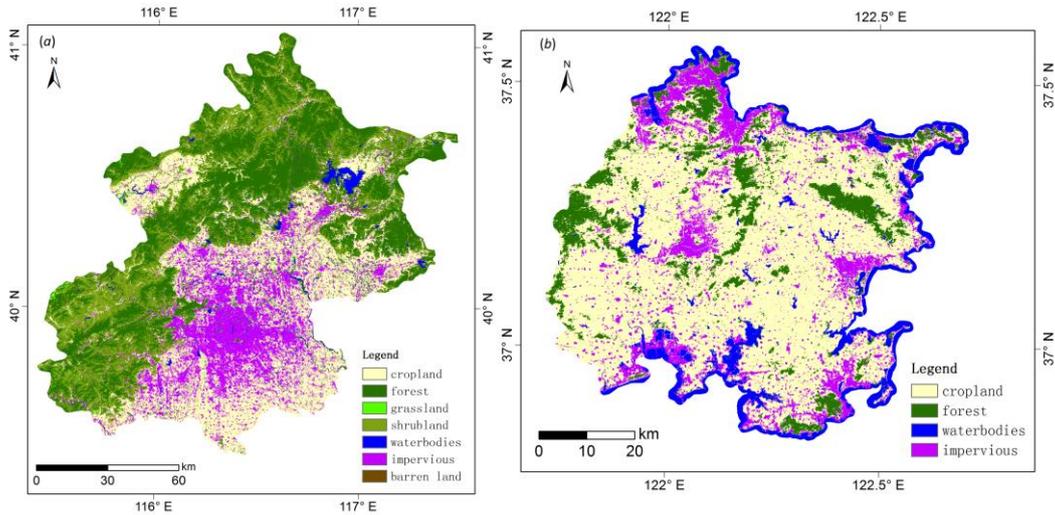

Figure 5. Land-cover mapping results obtained using the proposed method.

Quantitative comparisons were performed using our validation sample sets. Overall accuracies for the nine land-cover maps (MAP-A1, MAP-B1, MAP-LL, MAP-AM1, MAP-AB-SVM, MAP-AB-RF, MAP-R1, MAP-A2, and MAP-R2) are 74.0%, 74.0%, 81.1%, 78.4%, 78.4%, 80.6%, 81.8%, 83.0%, and 86.8%, respectively.

Tables 1-4 show confusion matrices for the four maps of Arear-1, MAP-A1, MAP-B1, MAP-LL, and MAP-R1. The confusion matrix for MAP-B1 was computed using all validation sample points, including those located in cloud and cloud shadow regions. An improvement can be seen from the result of the proposed method. The numbers of correctly classified sample units for almost all classes in MAP-LL and MAP-R1 are larger than those in MAP-A1 and MAP-B1. It is worth noting that information from MODIS data was used only in the cloud and cloud shadow regions of Landsat-B1. There are 40 validation sample points located in these regions. And 27, 19, 30, and 33 of them were correctly classified in MAP-A1, MAP-B1, MAP-LL, and MAP-R1,

respectively. Therefore, the overall accuracy for MAP-R1 is only a little higher than that for MAP-LL. However, the results can provide some indications about the advantage of the proposed method. All misclassified points in MAP-LL were misclassified in MAP-A1 and MAP-B1, and all misclassified points in MAP-R1 were misclassified in MAP-LL. This means that no new errors were introduced when combining multi-source information through the proposed method, at least for the validation points located in the cloud and cloud shadow regions. No improvement was obtained when combining information from MODIS data and the intermediate result *LL* in the cloud-free regions of Landsat-B1. On the contrary, information from MODIS data were used for the whole study area when acquiring MAP-AM1, and an improvement can be seen (overall accuracy, 78.4% versus 74.0%). Since only data quality was used as prior knowledge, information from MODIS data did not help much when multi-temporal Landsat data were used. More prior knowledge should be used for a further improvement.

Table 1. Confusion matrix for MAP-A1 (overall accuracy: 74.0%).

|  | Name | Ground reference data |  |  |  |  |  |  | Total | UA (%) |
|---|---|---|---|---|---|---|---|---|---|---|
|  |  | CR | FR | GR | SHR | WB | IMP | BL |  |  |
| Classification | CR | 93 | 4 | 5 | 6 | 0 | 8 | 0 | 116 | 80 |
| | FR | 9 | 97 | 0 | 21 | 0 | 0 | 0 | 127 | 76 |

| | | | | | | | | | |
|---|---|---|---|---|---|---|---|---|---|
| | GR | 2 | 0 | 1 | 1 | 0 | 0 | 0 | 4 | 25 |
| | SHR | 2 | 33 | 0 | 46 | 0 | 0 | 0 | 81 | 57 |
| | WB | 0 | 0 | 0 | 0 | 9 | 0 | 0 | 9 | 100 |
| | IMP | 8 | 0 | 3 | 0 | 0 | 75 | 2 | 88 | 85 |
| | BL | 3 | 1 | 0 | 4 | 2 | 0 | 4 | 14 | 29 |
| | Total | 117 | 135 | 9 | 78 | 11 | 83 | 6 | 439 | |
| | PA (%) | 79 | 72 | 11 | 59 | 82 | 90 | 67 | | 74.0 |

Note: CR, cropland; FR, forest; GR, grassland; SHR, shrubland; WB, waterbodies; IMP, impervious; BL, barren land; UA, user accuracy; PA, producer's accuracy.

Table 2. Confusion matrix for MAP-B1 (overall accuracy: 74.0%).

| | | Ground reference data | | | | | | | | |
|---|---|---|---|---|---|---|---|---|---|---|
| | Name | CR | FR | GR | SHR | WB | IMP | BL | Total | UA (%) |
| Classification | CR | 95 | 10 | 5 | 5 | 1 | 7 | 1 | 124 | 77 |
| | FR | 11 | 97 | 0 | 19 | 1 | 2 | 0 | 130 | 75 |
| | GR | 6 | 2 | 1 | 3 | 0 | 0 | 0 | 12 | 8 |
| | SHR | 3 | 25 | 1 | 47 | 0 | 0 | 0 | 76 | 62 |
| | WB | 0 | 1 | 0 | 0 | 9 | 0 | 0 | 10 | 90 |
| | IMP | 1 | 0 | 2 | 1 | 0 | 73 | 2 | 79 | 92 |
| | BL | 1 | 0 | 0 | 3 | 0 | 1 | 3 | 8 | 38 |
| | Total | 117 | 135 | 9 | 78 | 11 | 83 | 6 | 439 | 0 |

| | | PA (%) | 81 | 72 | 11 | 60 | 82 | 88 | 50 | | 74.0 |

Note: CR, cropland; FR, forest; GR, grassland; SHR, shrubland; WB, waterbodies; IMP, impervious; BL, barren land; UA, user accuracy; PA, producer's accuracy.

Table 3. Confusion matrix for MAP-LL (overall accuracy: 81.1%).

| | | Ground reference data | | | | | | | | |
|---|---|---|---|---|---|---|---|---|---|---|
| | Name | CR | FR | GR | SHR | WB | IMP | BL | Total | UA (%) |
| | CR | 103 | 6 | 6 | 4 | 0 | 7 | 0 | 126 | 82 |
| | FR | 7 | 110 | 0 | 19 | 0 | 0 | 0 | 136 | 81 |
| | GR | 0 | 0 | 2 | 2 | 0 | 0 | 0 | 4 | 50 |
| | SHR | 2 | 19 | 0 | 51 | 0 | 0 | 0 | 72 | 71 |
| | WB | 0 | 0 | 0 | 0 | 10 | 0 | 0 | 10 | 100 |
| | IMP | 4 | 0 | 1 | 0 | 0 | 76 | 2 | 83 | 92 |
| Classification | BL | 1 | 0 | 0 | 2 | 1 | 0 | 4 | 8 | 50 |
| | Total | 117 | 135 | 9 | 78 | 11 | 83 | 6 | 439 | |
| | PA (%) | 88 | 81 | 22 | 65 | 91 | 92 | 67 | | 81.1 |

Note: CR, cropland; FR, forest; GR, grassland; SHR, shrubland; WB, waterbodies; IMP, impervious; BL, barren land; UA, user accuracy; PA, producer's accuracy.

Table 4. Confusion matrix for MAP-R1 (overall accuracy: 81.8%).

|  | Ground reference data | | | | | | | | |
|---|---|---|---|---|---|---|---|---|---|
| Name | CR | FR | GR | SHR | WB | IMP | BL | Total | UA (%) |
| CR | 104 | 6 | 6 | 4 | 0 | 7 | 0 | 127 | 82 |
| FR | 7 | 111 | 0 | 19 | 0 | 0 | 0 | 137 | 81 |
| GR | 0 | 0 | 2 | 2 | 0 | 0 | 0 | 4 | 50 |
| SHR | 2 | 18 | 0 | 52 | 0 | 0 | 0 | 72 | 72 |
| WB | 0 | 0 | 0 | 0 | 10 | 0 | 0 | 10 | 100 |
| IMP | 3 | 0 | 1 | 0 | 0 | 76 | 2 | 82 | 93 |
| BL | 1 | 0 | 0 | 1 | 1 | 0 | 4 | 7 | 57 |
| Total | 117 | 135 | 9 | 78 | 11 | 83 | 6 | 439 | |
| PA (%) | 89 | 82 | 22 | 67 | 91 | 92 | 67 | | 81.8 |

(Left label: Classification)

Note: CR, cropland; FR, forest; GR, grassland; SHR, shrubland; WB, waterbodies; IMP, impervious; BL, barren land; UA, user accuracy; PA, producer's accuracy.

The CPDs based on data quality were aimed at using information from different data sources selectively. We created a map by performing a similar process as the acquisition of MAP-LL in the cloud and cloud shadow regions, and setting $f_n = 0$ for all pixels. In this map, 27 of the 40 sample points were correctly classified. Three misclassified forests points of MAP-A1 were corrected in this map, and at the same time, new errors were introduced. This comparison shows that using multi-source information selectively is an important advantage of the proposed method.

The overall accuracies for MAP-AB-SVM (78.4%) and MAP-AB-RF (80.6%) are slightly lower than that for MAP-LL (81.1%). The parameters for the SVM classifier were finely tuned by using grid search method. The overall accuracy was much lower (75.4%) when using the original parameters used for training on single-date Landsat data. In addition to higher accuracy, the proposed method is an appropriate option when facing the temporal inconsistency problem (different number and different acquisition dates of images in large area land cover mapping) as described in section 1. Initial probabilities could be estimated for each of the single-date images, with training samples collected from spatio-temporal neighbourhood scenes (Gong et al. 2013). And then the initial probabilities could be combined to get better results using the proposed PGM.

The data for Area-2 were also used to show the effectiveness of the proposed method. Table 5 and 6 show confusion matrices for the maps MAP-A2 and MAP-R2. It can be seen that the numbers of correctly classified samples for all classes for MAP-R2 are all larger than those for MAP-A, and a few new errors exist. The creation of MAP-R2 was performed in the same way as acquiring MAP-AM1. Information from MODIS data were used for the whole study area. And similar improvement (overall accuracy, 86.8% versus 83.0%) can be seen.

Table 5. Confusion matrix for MAP-A2 (overall accuracy: 83.0%).

| | Ground reference data | | |
|---|---|---|---|

|  | Name | CR | FR | WB | IMP | Total | UA (%) |
|---|---|---|---|---|---|---|---|
| Classification | CR | 179 | 36 | 6 | 17 | 238 | 75 |
| | FR | 2 | 67 | 1 | 0 | 70 | 96 |
| | WB | 0 | 0 | 52 | 0 | 52 | 100 |
| | IMP | 4 | 0 | 6 | 53 | 63 | 84 |
| | Total | 185 | 103 | 65 | 70 | 423 | |
| | PA (%) | 97 | 65 | 80 | 76 | | 83.0 |

Note: CR, cropland; FR, forest; WB, waterbodies; IMP, impervious; UA, user accuracy; PA, producer's accuracy.

Table 6. Confusion matrix for MAP-R2 (overall accuracy: 86.8%).

|  |  | Ground reference data | | | | | |
|---|---|---|---|---|---|---|---|
|  | Name | CR | FR | WB | IMP | Total | UA (%) |
| Classification | CR | 181 | 30 | 5 | 13 | 229 | 79 |
| | FR | 2 | 72 | 2 | 1 | 77 | 94 |
| | WB | 0 | 0 | 58 | 0 | 58 | 100 |
| | IMP | 2 | 1 | 0 | 56 | 59 | 95 |
| | Total | 185 | 103 | 65 | 70 | 423 | |
| | PA (%) | 98 | 70 | 89 | 80 | | 86.8 |

Note: CR, cropland; FR, forest; WB, waterbodies; IMP, impervious; UA, user accuracy; PA, producer's accuracy.

The validation samples were randomly distributed in the study areas. Therefore, the proportions of the validation samples are close to the real area coverage of their corresponding classes. Grassland, waterbodies and barren land are rare in Area-1, and their user and producer's accuracies are not statistically stable with only a few validation points. This is not the case for Area-2.

The proposed method can be extended to handle more data sources other than Landsat and MODIS data. Knowledge-based rules could be added to the extended PGM, and a larger graph structure is needed. CPDs could be computed using more information other than data quality, such as information learned from training samples.

## 5. Conclusion

Combining information from multiple data sources is an effective approach for high accuracy land-cover mapping. Methods based on pixel-level fusion are data specific. Useful information as well as noise will be mixed together if one simply concatenates features from multiple data sources.

The PGM approach proposed in this study is more flexible compared with data fusion methods. Independent classifiers were applied to the MODIS time series and to each single-date Landsat image. The initial probability estimation results were then fused using a PGM. Information from different sources was selectively used, so that useful information was obtained and few errors were introduced. Conclusions can be obtained by quantitative comparison. The original classification results for Landsat-A1 (MAP-A1, overall accuracy: 74.0%) and Landsat-B1 (MAP-B1, overall accuracy: 74.0%) were improved using the proposed PGM (MAP-LL, overall accuracy: 81.1%). The overall accuracy of the proposed method is 81.8% when adding MODIS data (MAP-R1). Information from MODIS data can help much when single-temporal Landsat data was used. MAP-AM1 for Area-1 (overall accuracy: 78.4% versus 74.0%) and MAP-R2 for Area-2 (overall accuracy: 86.8% versus 83.0%) show similar improvements by combining MODIS data and single-temporal Landsat data.

A comparison of results in local regions covered by cloud and cloud shadow was also performed. The number of validation points located in these regions is 40. 27 of these were correctly classified in MAP-A1, and 19 in MAP-B1. The number of correctly classified points is 30 in MAP-LL, and 33 in MAP-R1. No new errors were introduced when applying the proposed PGM. The number of correctly classified points was 27 in the combined result when information from Landsat-B1 was not used selectively. Both improvement and deterioration occurred.

The proposed method can be extended for large-area land-cover mapping with temporally inconsistent multi-temporal Landsat images (different number and different acquisition dates of images in large area land cover mapping, as described in section 1). Multiple data sources as well as knowledge-based rules can be added to create a large graph structure for an extended PGM.

**Disclosure statement**

No potential conflict of interest was reported by the authors.

**Funding**

This research was partially supported by a Meteorology Scientific Research in the Public Welfare [GYHY201506010] a National High Technology Grant from China [2013AA122804], the National Natural Science Foundation of China [41001274], the National Natural Science Foundation of China [41271423] and the Open Fund of State Key Laboratory of Remote Sensing Science [OFSLRSS201514].